# Ab-initio simulations on growth and interface properties of epitaxial oxides on silicon

C.J. Först[1,2,*], C.R. Ashman[1,†], K. Schwarz[2], P.E. Blöchl[1]

[1] *Clausthal University of Technology, Institute for Theoretical Physics, Leibnitzstr. 10, 38678 Clausthal-Zellerfeld, Germany;*
[2] *Vienna University of Technology, Institute for Materials Chemistry, Getreidemarkt 9/165-TC, A-1060 Vienna, Austria;*
[*] *present address: Massachusetts Institute of Technology, Department of Nuclear Science and Technology, Massachusetts Avenue 77, Cambridge, MA 02139, USA*
[†] *present address: Materials Science and Technology Division, Naval Research Laboratory, Washington D.C., 20375, USA*
Tel: +49-5323-722021      email: peter.bloechl@tu-clausthal.de

**Abstract**

The replacement of $SiO_2$ by so-called high-k oxides is one of the major challenges for the semiconductor industry to date. Based on electronic structure calculations and ab-initio molecular dynamics simulations, we are able to provide a consistent picture of the growth process of a class of epitaxial oxides around SrO and $SrTiO_3$. The detailed understanding of the interfacial binding principles has also allowed us to propose a way to engineer the band-offsets between the oxide and the silicon substrate.

## 1. Introduction

The replacement of $SiO_2$ as a gate dielectric in microelectronic devices is one of the key challenges of the semiconductor industry in the next years [1]. Due to their larger dielectric constants, so-called high-k oxides can be deposited with greater physical thicknesses which reduces the quantum-mechanical leakage currents through insulating oxide layers. The growth of these, mostly transition metal containing oxides on the technologically relevant Si(001) surface has proven to be highly challenging. While in a first phase amorphous oxides on the basis of $ZrO_2$ and $HfO_2$ in connection with an ultra-thin $SiO_2$ layer will be employed, industry requires solutions for crystalline oxides with an epitaxial interface to the silicon substrate as soon as 2013 [1]. So far, however, there is only one clear demonstration of an atomically well defined interface between silicon and a high-k oxide, namely $SrTiO_3$. Despite the clear experimental evidence of an atomically well defined interface [2], the interfacial stoichiometry and structure remained elusive. Key process parameters as well as the electronic properties at the interface are still under debate. Using ab-initio molecular dynamics simulations we have been able to unravel the growth process of $SrTiO_3$ on Si(001) and propose a way to engineer the band offsets with respect to silicon [3,4]. The detailed understanding of the chemistry of group II to IVa transition metals on the silicon (001) surface has proven to be vital in order to rationalize the mechanisms of interface formation. Recent results show that the basic binding principles can also be applied to investigate the interfacial chemistry between silicon and $LaAlO_3$ [5]. This paper summarizes the results obtained by the authors in this context. In depth information can be found in our previous publications [3,4,6,7]. Other theoretical studies are [8–10].

## 2. Metal adsorption on Si(001)

If one wants to maintain the integrity of the substrate, it is important to avoid the oxidation of silicon. Therefore it is mandatory to start with the deposition of the metal. The critical part during growth of an oxide with molecular beam epitaxy is the deposition of the metal and the oxidation of the adsorbed metal layers. Therefore a detailed understanding of the adsorption of the metal layer is important to guide the growth process. We have investigated the deposition of metal ions selected from the three most relevant groups in the context of high-k oxides on the silicon (001) surface. These are the divalent earth-alkali metals and the three- and four-valent transition metals exemplified by strontium (Sr) [4], lanthanum (La) [7], and zirconium (Zr) [6].

We simulated a wide range of adsorbed metal layers. Scanning the relevant phase space is a considerable task. The search is guided by chemical insight, geometric classification and ab-initio simulations that provide an unbiased exploration of a small region of the phase space. While this approach cannot exclude with certainty that a particular low-energy structure has been missed, using a mixture of various search strategies and a sufficiently wide search space seems to yield fairly reliable results.

Figure 1 shows the results for the example of Sr adsorption on silicon. The symbols indicate the adsorption energies per (1 × 1) silicon surface cell as function of coverage in monolayers (represented by appropriate supercells). The adsorption energies are obtained relative to a reservoir of bulk silicon and the most stable silicide, in this case $SrSi_2$. Only the lower envelope is physically relevant. At those coverages, where the line segments meet, we predict stable phases. The straight line segments indicate coexistence of the two adjacent stable phases. As the coverage increases within one of the line segments, the surface area of one phase is converted into that of the other phases. Thus we are able to predict the sequence of stable phases at zero temperature. From simple thermodynamic considerations we expect that the qualitative features will not change at finite, but not too high temperatures.

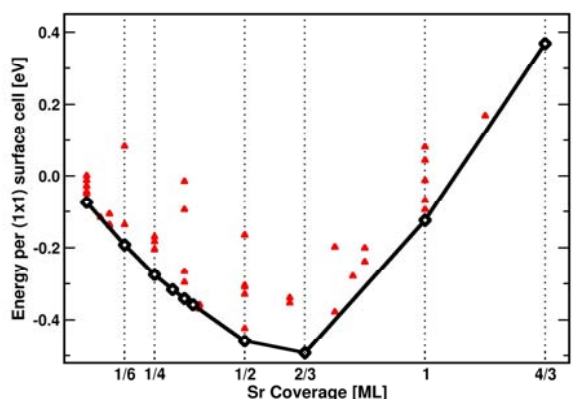

*Fig.1 Surface energy versus coverage for Sr. The open diamonds represent thermodynamically accessible structures, the triangles correspond to metastable structures.*

One of the main problems during growth of high-k oxides is the formation of silicide grains. Bulk thermodynamic stability of the relevant metals has been investigated by Schlom [11,12] but may be misleading for film growth in the monolayer range, since it neglects the effects of epitaxial strain and does not account for the binding of the metal layer to the substrate. Going one step further is to compare the energies of the adsorbed metal layer with that of a bulk silicide. Since the adsorption energies have been calculated with respect to the silicide as particle reservoir, we obtain the stability limit, where the lower envelope in figures 1 and 2 changes its slope to positive values.

In the case of Sr we calculate coverages up to 2/3 monolayer (ML) as thermodynamically stable [4], whereas for the La surface we find reconstructions only to be stable up to the coverage of 1/3 ML [7] (compare figure 2). In case of Zr and Hf, all surface reconstructions are unstable with respect to silicide formation, which excludes the possibility of metal pre-deposition [6]. In the case of

Zr we find that the adsorption structures become more stable with increasing coverage, indicating that the metal atoms agglomerate spontaneously. In addition we observed the onset of silicide formation, when Zr atoms go subsurface. Zr-silicide formation has also been identified experimentally [13].

The stable surface reconstructions found for Sr and La are driven by the atomic and electronic topology of the Si(001) surface. The basic principle will shortly be summarized with the example of Sr. La behaves conceptually similar at low coverages and, due to a change in oxidation state from 3+ to 2+ above 1/3 ML, even identical to Sr at high coverages [7].

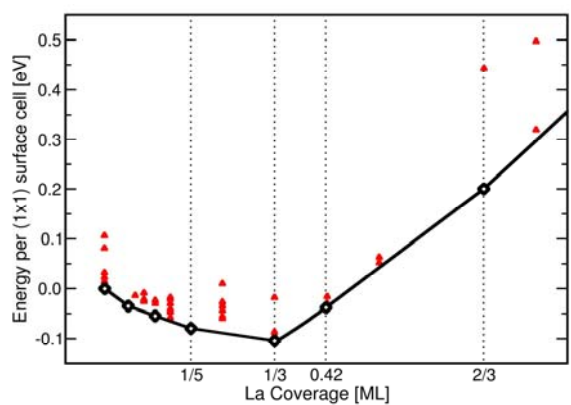

Fig. 2 Surface energy versus coverage for La. The open diamonds represent thermodynamically accessible structures, the triangles correspond to metastable structures.

The chemistry of Si and Sr is probably best understood in terms of the Zintl-Klemm concept [14–16]. It explains in a simple ionic picture the relationship between stoichiometry and structure for a wide range of compounds between electronegative elements, essentially groups IV to VII, and electropositive elements, mainly groups I and II. The electropositive element, for example Sr, donates its electrons to the electronegative element, such as Si. Each electron of the electropositive element can saturate one dangling bond of the electronegative element. As a result the electronegative element forms structures which have a smaller number of covalent bonds. $Si^0$ forms four bonds, $Si^{1-}$ forms three bonds, $Si^{2-}$ forms two bonds, and so on. The bonding is as if the atom would be shifted in the periodic table by one row to the right for each electron it receives.

On the Si(001) surface the electrons form silicon dimers, where each atom has three covalent bonds and one dangling bond. The two unpaired electrons in the silicon dangling bonds pair up in one of the silicon atoms. This results in the buckled dimer reconstruction, where the upper atom is a negative Si-ion and the lower atom is a positive Si-ion, because a 5-electron species such as $Si^-$ favors $sp^3$-hybridization, while a 3-electron species such as $Si^+$ favors a trigonal-planar coordination.

Translating the Zintl-Klemm concept to the silicon (001) surface implies that each Sr ad-atom will saturate the dangling bonds of one silicon dimer with its two valence electrons. A saturated dimer looses its buckling, since all dangling bonds are filled and both Si atoms prefer the tetrahedral $sp^3$ configuration. The atomic structure around an isolated Sr-ad-atom is schematically shown in figure 3. The preferred adsorption site is in the center of four dimers. One of the neighboring Si dimers is unbuckled due to the electron donation from the ad-atom. The other three dimers are orientated such that the upper, and thus negatively charged Si atom points towards the $Sr^{2+}$ ion. The energy penalty for placing a lower and thus positively charged silicon atom next to a Sr ad-atom is roughly 0.4eV. Therefore the ad-atoms pin the dimer buckling as observed experimentally [17].

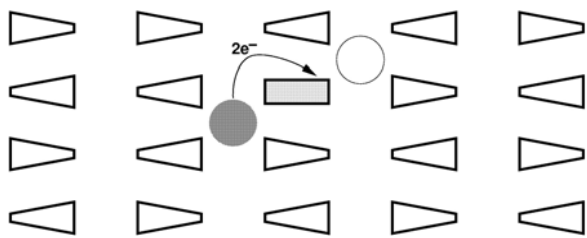

Fig. 3 Schematic representation of the isolated Sr ad-atom at the preferred adsorption site position. The filled circle represents the Sr ad-atom, the rectangle represents a filled and therefore unbuckled Si dimer. The triangles represent buckled dimers. The flat side of a buckled dimer indicates the upper Si atom with a filled dangling bond, whereas the pointed side indicates the lower Si atom with the empty dangling bond. The charge transfer from the Sr ad-atom to one of the surrounding dimers is indicated by the arrow, the preferred adsorption site in the neighboring valley by the open circle.

The filled dimer offers a preferred adsorption site in the next valley as indicated by the open circle in figure 3. As a result, diagonal and zig-zag chain structures turn out to be the thermodynamically stable reconstructions at low coverages. At 1/6 ML these chains condense (figure 4). It is not possible to stack them with a separation of only two instead of three lattice constants because that would imply that a lower and thus positively charged Si atom points towards a Sr ion which involves a large energy penalty as dicussed above. This is the reason for a distinct phase at a coverage of 1/6th ML. The next stable phase is made of double chains at 1/4 ML as shown in figure 4.

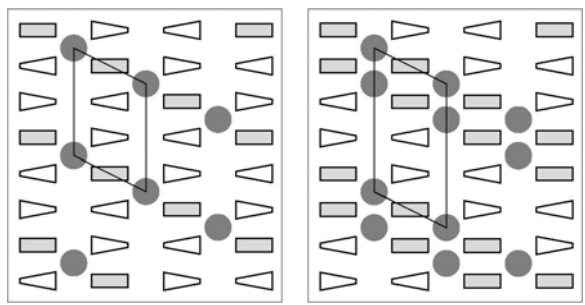

*Fig. 4 Chain structures of Sr ad-atoms at low coverages. The symbols are explained in figure 3. The surface unit cells are outlined. Left: Condensed diagonal chains of Sr at a coverage at 1/6 ML. Right: Diagonal double chains of Sr at a coverage of 1/4 ML.*

The phases at 1/6 ML and at 1/4 ML correspond to those observed by McKee et al. by their $(1 \times 3)$ and $(1 \times 2)$ RHEED pattern [18]. The presence of these phases has been one main argument for proposing a solid-state transformation to a silicide phase as expected from the bulk phase diagram. Our calculations, however, clearly show that the structural model proposed by McKee [2] is higher in energy than the structures discussed above. Therefore, we rule out the model of a transformation into a silicide layer of the form proposed by McKee et al [2].

At one half monolayer, the Sr ad-atoms occupy all favorable positions in the center of four dimers. We labeled this coverage as the "canonical coverage" because all dangling bond states are saturated. Consequently we find that there are no surface states deep in the gap of silicon. This surface is isoelectronic to a hydrogen terminated silicon surface and thus is expected to be chemically comparably inert. Experimentally it has been found that the surface is exceptionally inert against oxidation at 1/2 ML of Sr [19].

Above 1/2 ML, the electrons donated by Sr ad-atoms enter the dimer anti-bonding states leading to a partial breakup of the dimer bonds. At 2/3 ML we observe a $(3 \times 1)$ reconstruction with alternating rows of Si dimers and isolated Si atoms.

Lanthanum adsorption follows similar principles with two main differences [7]. Firstly, La donates three electrons instead of two. This changes the phase-diagram considerably, despite very similar building principles. Secondly, La deviates from the Zintl-Klemm concept in that it changes its charge state from being a formal $La^{3+}$ ion at lower coverages to a $La^{2+}$ ion at coverages above 1/3 ML. However, the structures above 1/3 ML are no more thermodyamically stable against the formation of La silicide.

Our studies of the adsorption of two-, three-, and four-valent metals on Si(001) led to a unified picture of the processes and has added new insight to a number of experimental results.

## 3. Interface of $SrTiO_3$ and Si(001)

The chemical bonding in silicon and $SrTiO_3$ is fundamentally different. While silicon is a covalently bonded material, $SrTiO_3$ is an ionic crystal with some covalent character in the Ti–O bonds. More specifically, $SrTiO_3$ crystallizes in the perovskite structure with Ti being octahedrally coordinated by oxygen and Sr placed in a cubic oxygen cage. The (001) planes of $SrTiO_3$ are alternating SrO and $TiO_2$ planes and thus electronically saturated, that is unable to form covalent bonds. The SrO terminated surface does not exhibit states in the band gap. An electronically saturated Si–$SrTiO_3$ stack must thus exhibit an interfacial layer which provides a covalent bonding environment towards the silicon substrate and in addition an ionic template compatible with that of $SrTiO_3$. The only Sr covered surface meeting these requirements is the reconstruction at 1/2 ML. It is the only one which saturates all silicon dangling bonds and does not have surface states in the band-gap of silicon. The quasi-ionic interaction of Sr with Si furthermore prepares an ionic template with a formal charge distribution as visualized in figure 5. The resulting two-dimensional ionic layer is compatible with the NaCl–type charge pattern of a

SrO–terminated SrTiO$_3$ crystal. Using molecular dynamics we simulated the "deposition" of SrTiO$_3$ on top of this template. The resulting structure is shown in figure 6.

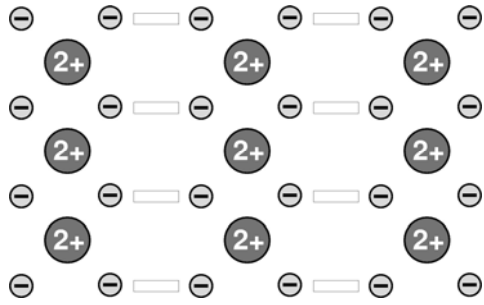

*Fig. 5 Charge pattern of the silicon surface covered by 1/2 ML of Sr. The Sr ions have a formal charge of 2+, Si of 1− due to the filled dangling bond. The empty rectangles denote the dimer bonds, included to orient this figure with the lower left panel of figure 6.*

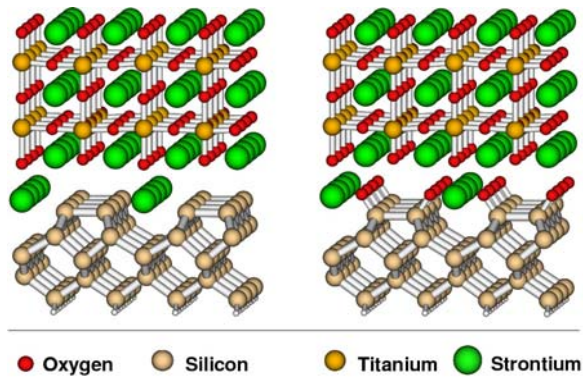

*Fig. 6 The two relevant interface structures between silicon and SrTiO$_3$.*

### 4. Band offset engineering

An electrically inactive interface is only one of the requirements a high-k oxide has to meet. A common problem with high-k oxides is a conduction-band offset between silicon and the oxide which is too small. This offset acts as an electron injection barrier, which prevents electrons from being drawn into the gate from the channel via the gate-oxide. For device applications, the conduction and valence band offsets have to be in the range of 1eV or larger. However, for the interface just introduced (left panel of figure 6) we calculate a conduction band offset of only 0.2eV, which is in line with measurements done on related interfaces.

Depending on growth and annealing conditions, the oxygen content of the interface can change. Oxygen can be introduced during growth, or it can diffuse in from the oxide. Therefore we investigated the stability of the interface against oxidation [3]. We find that oxygen first attacks the filled dangling bonds of the silicon dimer atoms. In this way exactly one monolayer of oxygen can be selectively introduced at the interface. The resulting interface structure is shown in the right panel of figure 6. We confirmed that this interface can be formed without growing an interfacial SiO$_2$ layer by oxidizing the substrate. Interestingly we find that this oxygen monolayer at the interface creates a large charge dipole, which increases the conduction band offset by 1.1eV. The resulting injection barrier of 1.3eV is in line with technological requirements. The origin of this charge dipole is the oxidation of the interface silicon atoms from Si$^-$ to Si$^+$ when the O$^{2-}$ ions are introduced.

To the best of our knowledge, SrTiO$_3$ is the only high-k oxide, for which a controlled growth of an atomically well defined interface with silicon has been demonstrated. Originally this system has nearly been discarded from the list of potential high-k oxides because of the small electron injection barrier. Our finding of adjusting the band offset by selective oxidation of the interface bears the hope that SrTiO$_3$ can still be made useful for device applications.

### 5. Conclusions

We provided a detailed picture of metal adsorption from groups II to IVa on silicon (001). We have rationalized the sequence of phases seen during metal adsorption in a simple, intuitive and generalizable picture.

Our calculations led to a structure model for the interface of SrTiO$_3$ and Si(001), which differs from earlier proposals. We furthermore investigated the chemical changes of the interface upon oxidation. We have shown that the electron injection barrier can be dramatically increased by controlled oxidation of the interface, in order to match the technological requirements.

**Acknowledgements**

We thank A. Dimoulas, J. Fompeyrine, J.-P. Loquet for useful discussions. This work has been funded by the European Commission in the project "INVEST" (Integration of Very High-K Dielectrics with CMOS Technology), the European project ET4US "Epitaxial Technologies for Ultimate Scaling" and by the AURORA project of the Austrian Science Fond. Parts of the calculations have been performed on the Computers of the "Norddeutscher Verbund für Hoch- und Höchstleistungsrechnen (HLRN)". This work has benefited from the collaborations within the ESF Programme on 'Electronic Structure Calculations for Elucidating the Complex Atomistic Behavior of Solids and Surfaces'.